\documentclass[%
 reprint,
 amsmath,amssymb,
 aps,
 prx,
reprint,
]{revtex4-2}

\usepackage{graphicx}
\usepackage{dcolumn}
\usepackage{bm}
\usepackage{hyperref}

\usepackage{siunitx}
\usepackage{booktabs}
\usepackage{braket}

\begin{document}

\preprint{APS/123-QED}

\title{On the Relationship Between Antibunching and Entanglement in Resonance Fluorescence}

\author{Xin-Xin Hu}
\author{Gabriele Maron}
\author{Luke Masters}
\author{Arno Rauschenbeutel}
\email{arno.rauschenbeutel@hu-berlin.de}
\author{J\"urgen Volz}
\email{juergen.volz@hu-berlin.de}

\affiliation{Department of Physics, Humboldt Universit\"at zu Berlin, 10099 Berlin, Germany}

\date{\today}

\begin{abstract}

Photon antibunching in resonance fluorescence---the emission from a single, resonantly driven two-level quantum emitter---is a paradigmatic signature of nonclassical light. Photon entanglement, by contrast, manifests as correlations that can defy any classical description and is typically regarded as a distinct quantum effect. Here, we experimentally extract pairs of narrowband, time-bin–entangled photons from the antibunched resonance fluorescence of a single trapped atom. We verify entanglement via violation of the CHSH Bell inequality and by reconstructing the two-photon density matrix. The observed correlations vanish when the coincidence time window exceeds the antibunching timescale, revealing underlying multimode entanglement in the emitted field. Our results establish a direct link between photon antibunching and photon–photon entanglement, unifying two canonical signatures of nonclassical light.
\end{abstract}

\maketitle



\par Individually controlled two-level quantum systems can be realized using, e.g., laser-cooled atoms \cite{Phillips1998,Schlosser2001,Endres2016,Barredo2018,Bloch2008,Kaufman2014} or ions \cite{Monroe2013,Haffener2008}, quantum dots \cite{Kloeffel2013,Fattal2004}, single molecules \cite{Anderegg2019,Toninelli2021,Trebbia2010,Rezai2019} or defect centers \cite{Awschalom2018,DeLeon2021,Schroeder2017}. They constitute an exceptional platform for applications in quantum information and communication as well as fundamental studies of quantum optics and quantum mechanics. Their well-defined energy structure, whose interaction with light is a fundamentally quantum mechanical phenomenon, underlies many emerging quantum information technologies, particularly for the realization of quantum light sources \cite{Aharonovich2016,Somaschi2016}. 

\begin{figure}[ht]
\centering
\includegraphics[width=0.4\textwidth,keepaspectratio]{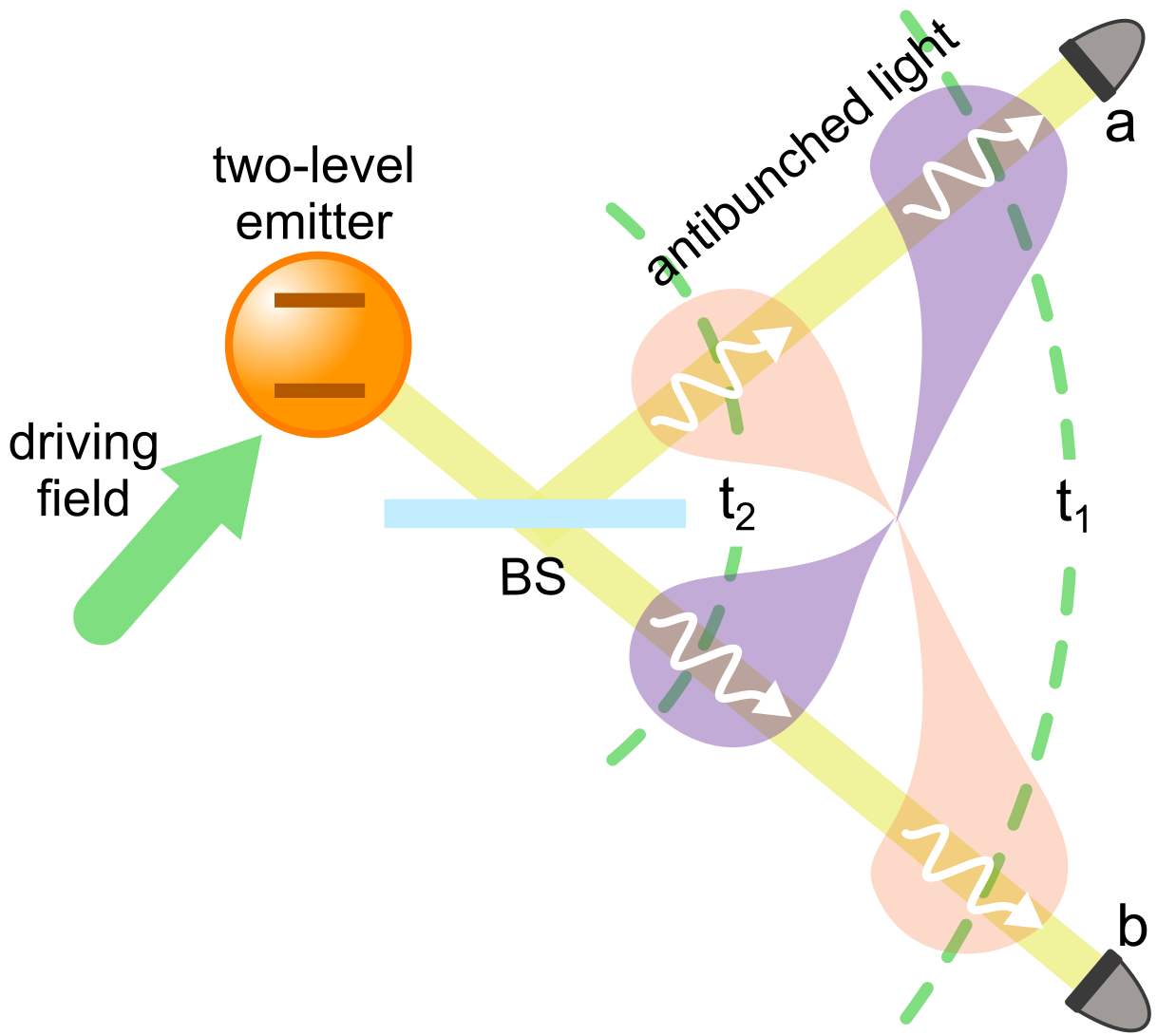}
\caption{\label{fig1ext} Resonance‐fluorescence–mediated entanglement. 
A single two-level quantum emitter, continuously driven by a coherent driving field, scatters light that features antibunched photon statistics, i.e., each emitted photon inhabits a temporally-distinct time-bin. Sending the emitted light onto a beamsplitter (BS), this absence of simultaneous photon emission ensures that one never observes a photon coincidence detection in the two output modes, $a$ and $b$ of the beamsplitter. When there is a photon in each output mode, the emitted light is then in the maximally entangled superposition state of: early emission into mode $a$ and late emission into mode $b$ (purple shaded), plus early emission into mode $b$ and late emission into mode $a$ (orange shaded).   
}
\end{figure}

\par When coherently driven by near-resonant light, such two-level emitters scatter photons, giving rise to so-called resonance fluorescence. The corresponding stream of photons displays hallmark quantum features such as photon antibunching, evidenced by a vanishing second-order correlation function at zero time-delay \cite{Kimble1977}. Although photon antibunching has been observed in many experiments, recent research efforts have been put into understanding and harnessing the physical process underlying its occurrence in resonance fluorescence \cite{Valle2012,Phillips2020,hanschke2020}. In this context, it has been experimentally shown that antibunching arises from destructive quantum interference between the two-photon components of the so-called coherent and incoherent scattering processes. Recently, pioneering works have experimentally shown that when removing the incoherent or the coherent component via spectral filtering, the remaining light field loses all photon correlations \cite{hanschke2020}, or consists predominantly of energy-time entangled photon pairs \cite{Masters2023,Liu2024,Wang2025}, respectively.

\begin{figure*}[ht]
\centering
\includegraphics[width=1.0\textwidth,keepaspectratio]{fig1_v9.png}
\caption{\label{fig1} Experimental set-up.
(a) A single two-level atom is loaded from a MOT into an optical dipole trap, and resonantly driven by an external laser with a residual detuning of $\Delta = 2\pi \times (2.56 \pm 0.16)$ MHz. Fluorescence photons from the trapped atom are collected using a high numerical aperture lens (${\textrm{NA}=0.55}$) and coupled into a single-mode fiber. (b) Second-order correlation functions ($g^{(2)}(\tau)$) of the collected fluorescence measured under different driving regimes. Clear antibunching is evident in the photon statistics in both the low and high driving regimes (respectively green and orange). The dark solid curves are theoretical fits, see Supplementary Materials. The shaded region indicates the variable-width coincidence window, $\delta t$, used for the Bell-inequality violation. (c) A 50:50 beamsplitter (BS) equally splits up the collected fluorescence and sends it into an all-fiber-based Franson-type setup, consisting of an unbalanced Mach-Zehnder interferometer on \emph{Alice's} (orange) and \emph{Bob's} (purple) side, respectively. Each of these has a long ($l$) and short ($s$) arm, with a delay time of $\Delta t_{A(B)}$. The path length difference is stabilized and set to impart a relative phase $\phi_{A(B)}$. Photons are detected in the output of each interferometer using superconducting nanowire single-photon detectors (SNSPDs), and their coincidence detection results in a maximally entangled Bell state, $\ket{\Psi_\textrm{Bell}}$.}
\end{figure*}

\par In this Letter, we elucidate the relation between antibunching and entanglement in {\it unfiltered} resonance fluorescence that comprises the full Mollow spectrum. For this, we simply send the collected fluorescence of a single trapped atom onto a 50:50 beamsplitter, thereby generating two distinct spatial modes, which feature strong temporal correlations due to the photon antibunching initially present. In particular, thanks to these correlations, a time-bin coincidence between the two spatial modes projects their state onto a maximally entangled Bell state. We verify this entanglement by violating a Clauser–Horne–Shimony–Holt (CHSH) type Bell inequality \cite{CHSH69} as well as by reconstructing the density matrix through quantum state tomography. Our experimental results demonstrate a quantitative link between antibunching and entanglement---the two fundamental non-classical properties of light that result in violations of Cauchy's and Bell's inequalities, respectively. 


\par To understand the origin of the entanglement, one has to consider that, light emitted by a single two-level quantum emitter under continuous and near-resonant coherent excitation exhibits photon antibunching. This means that at any given time of detection and distance from the atom, the fluorescence light contains, at most, a single photon. We now consider the situation in Fig.~\ref{fig1ext}, where the fluorescence is sent onto a 50:50 beamsplitter. We refer to the beamsplitter outputs as modes $a$ and $b$, respectively. We then assume that the four spatio-temporal modes described by the operators, $a^{\dagger}_{t_1}b^{\dagger}_{t_1}$, $a^{\dagger}_{t_2}b^{\dagger}_{t_2}$, $a^{\dagger}_{t_1}b^{\dagger}_{t_2}$ and $a^{\dagger}_{t_2}b^{\dagger}_{t_1}$ contain in total two photons. Here, $a^{\dagger}_{t_{1}}$ ($a^{\dagger}_{t_{2}}$) and $b^{\dagger}_{t_{1}}$ ($b^{\dagger}_{t_{2}}$) are the creation operators of a photon in mode $a$ and $b$ at time $t_1$ and $t_2$. In this case, the light is described by the state 
\begin{equation}
     |\Psi\rangle \propto \psi(\tau)\left[ a_{t_1}^\dagger b_{t_2}^\dagger + a_{t_2}^\dagger b_{t_1}^\dagger \right]\ket{0}+\psi(0)\left[ a_{t_1}^\dagger b_{t_1}^\dagger + a_{t_2}^\dagger b_{t_2}^\dagger \right]\ket{0}.
     \label{eq:entangled1}
\end{equation}
Here, $\psi(\tau)$ and $\psi(0)$ are the amplitudes of having two fluorescence photons featuring a time delay $\tau=t_1-t_2$ and at zero time delay, respectively, see Supplementary Materials. In the case of uncorrelated light, $\psi(\tau)=\psi(0)$ and Eq.~(\ref{eq:entangled1}) corresponds to a separable state. For perfectly antibunched light, however, we have $|\psi(0)|^2=0$ and Eq.~(\ref{eq:entangled1}) corresponds to a maximally entangled Bell state. Remarkably, in the absence of external decoherence mechanisms, the entangled state described by Eq.~(\ref{eq:entangled1}) is valid for any time difference $\tau\neq0$ and persists regardless of the driving strength. This is because the fundamental quantum interference that underlies photon antibunching ensures that the two-photon probability $\left|\psi(\tau=0)\right|^2=0$, prevails for any excitation strength, i.e., from weak excitation through to large saturation, see Supplementary Materials.


\par To experimentally test the entanglement, we use a single trapped ${^\text{85}}$Rb atom (Fig.~\ref{fig1}(a)) that produces antibunched fluorescence light (Fig.~\ref{fig1}(b)) and send it onto a beamsplitter that routes the photons to \emph{Alice} and \emph{Bob}, respectively. We transform the state in Eq.~(\ref{eq:entangled1}) into a readily experimentally detectable quantum state by employing a Franson-type interferometer setup (Fig.~\ref{fig1}(c)). Each observer, \emph{Alice} and \emph{Bob}, uses an unbalanced Mach-Zehnder interferometer that exhibits a path delay $\Delta t_{A(B)} \approx 46$~\unit{ns}, longer than the correlation time $\tau=(2\gamma)^{-1}$ of the antibunching, where $\tau=26.5$~\unit{ns} is the excited state lifetime of $^{85}$Rb \cite{steckDline}. For photons that fulfill $|t_1-t_2|=\Delta t$ (with $\Delta t \approx \Delta t_{A(B)}$), the entangled state in Eq.~(\ref{eq:entangled1}) is thus transformed into the two-mode entangled state
\begin{equation}
    |\Psi_\textrm{Bell}\rangle =\frac{1}{\sqrt{2}}\left[ a_s^\dagger (t)b_l^\dagger (t) + a_l^\dagger (t)b_s^\dagger (t)\right]\ket{0}, 
    \label{eq:entangled2}
\end{equation}
where the path delay $\Delta t$ sets the particular emission time difference $t_1-t_2$ that we study. Here $a_s^\dagger$ ($b_s^\dagger$) and $a_l^\dagger$ ($b_l^\dagger$) are the creation operators of a photon in the short and long arm of \emph{Alice's} (\emph{Bob's}) interferometer, respectively. Equation~(\ref{eq:entangled2}) describes a maximally entangled Bell state. Its origin can intuitively be understood when considering that photons entering the Franson-type interferometer setup can take four possible path combinations, i.e., long–short $|l,s\rangle$, short–short $|s,s\rangle$, long–long $|l,l\rangle$, and short–long $|s,l\rangle$. However, photon antibunching ensures a temporal separation between the photons. This timing constraint suppresses the amplitudes for the states $|s, s\rangle$ and $|l, l\rangle$ and subsequently maps the initial entangled state in Eq.~(\ref{eq:entangled1}) to Eq.~(\ref{eq:entangled2}). 

\par The Franson interferometer setup also allows a direct analysis of the entangled state in different measurement bases. For this purpose, we place phase shifters in the long arms of \emph{Alice's} and \emph{Bob's} interferometers, thereby setting the respective phases $\phi_A$ and $\phi_B$. Recombining the modes at the output beamsplitters, the photons are routed into different output ports for which we assign a value of $+1$ to detections at port 1 ($a_1$ or $b_1$) and $-1$ to detections at port 2 ($a_2$ or $b_2$). In this way we can measure the expectation values $\langle \sigma_{\phi_A}\rangle$ and $\langle \sigma_{\phi_B}\rangle$ for the photons on \emph{Alice's} and \emph{Bob's} side, respectively, where $\sigma_{\phi} = \cos\phi\,\sigma_x + \sin\phi\,\sigma_y$, $\sigma_x$ and $\sigma_y$ are the Pauli matrices defined for the two-dimensional Hilbert space of the photon.


\begin{figure}[t]
\centering
\includegraphics[width=0.5\textwidth,keepaspectratio]{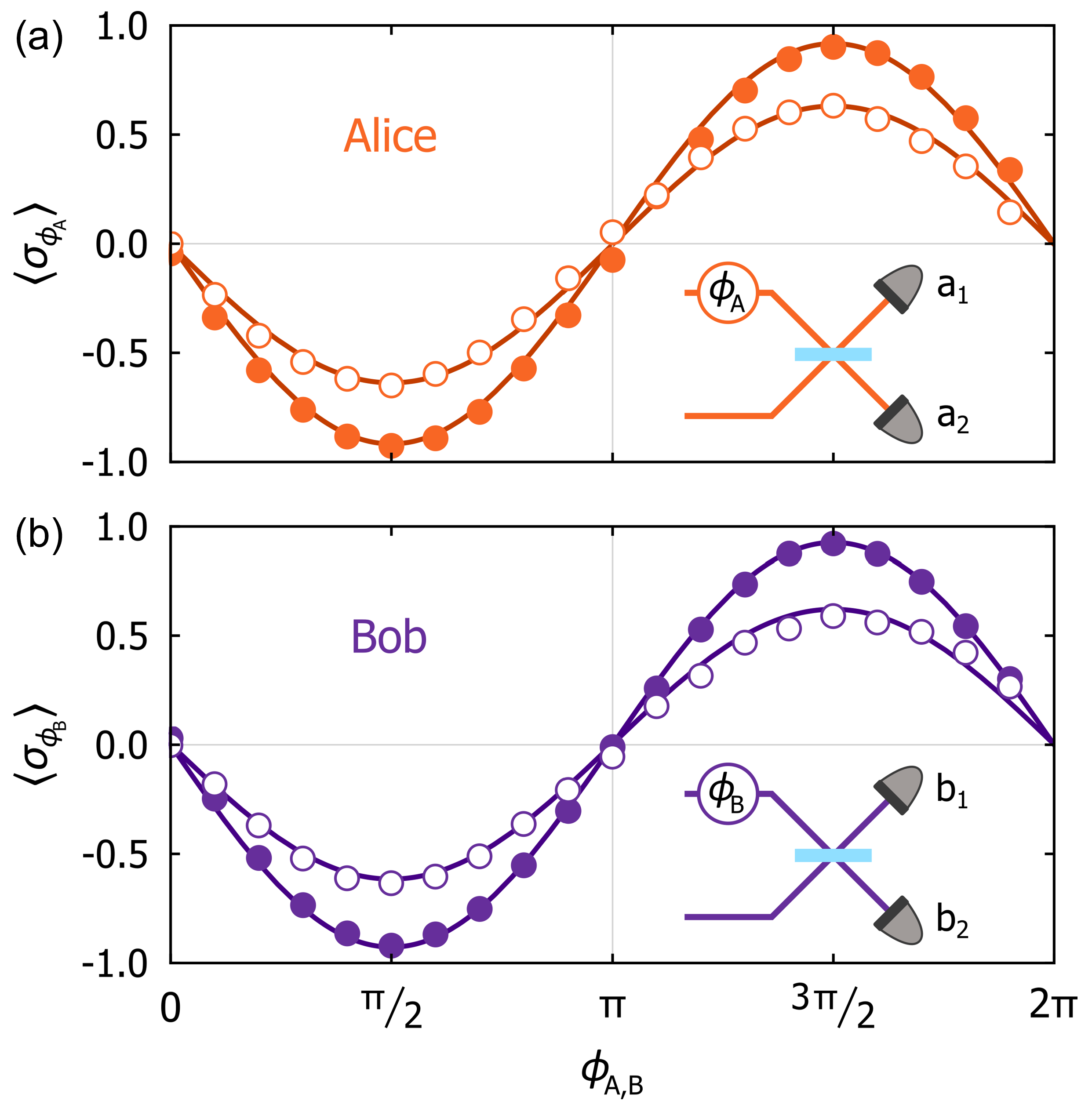}
\caption{\label{fig2} Single photon visibility. Measured expectation values, $\langle\sigma_{\phi_A}\rangle$ and $\langle\sigma_{\phi_B}\rangle$, obtained when launching resonance fluorescence through (a) \emph{Alice's} and (b) \emph{Bob's} interferometer for different phases $\phi_A$ and $\phi_B$, respectively. Sinusoidal fits (solid lines) to the data yield visibilities of $\textrm{V}_A=92\pm1\,\%$ and $\textrm{V}_B=93\pm1\,\%$ for low saturation (solid circles), and $\textrm{V}_A=61.8\pm1\,\%$ and $\textrm{V}_B=63.4\pm1\,\%$ for high saturation (open circles). The $1\sigma$ error bars are smaller than the displayed data points.}
\end{figure}

\par Scanning $\phi_A$ and $\phi_B$ and measuring the single photon expectation values $\langle \sigma_{\phi_A}\rangle$ and $\langle \sigma_{\phi_B}\rangle$ unveil single-photon interference fringes (Fig.~\ref{fig2}) with visibilities that depend on the driving strength. Under weak excitation (saturation parameter $s_0=0.10$), we observe a visibility of $\text{V}_\text{A} = 92\pm1\,\%$ and $\text{V}_\text{B} = 93\pm1\,\%$, while driving the atom at high saturation intensity ($s_0=2.75$) leads to a decrease of the visibilities to $61.8 \pm 1\,\%$ and $63.4 \pm 1\,\%$. This behavior originates from the dual composition of resonance fluorescence, that has driving-dependent contributions of the coherent and incoherent scattering components \cite{CohenTan2019}. Only the coherent component, that results from elastic photon-atom interactions and preserves the narrow linewidth of the excitation laser, exhibits a well-defined phase and thus high-contrast interference fringes. In comparison, the incoherently scattered component exhibits a broader spectral linewidth ($\delta\omega \geq 2\gamma$) and thus gives rise to a reduced interference contrast for an interferometric delay $\Delta t>(2\gamma)^{-1}$ as in our case. As the ratio of coherent to incoherent scattering decreases with increasing saturation, the interference visibility correspondingly diminishes, see Supplementary Materials for more details.

\begin{figure}[t]
\centering
\includegraphics[width=0.45\textwidth,keepaspectratio]{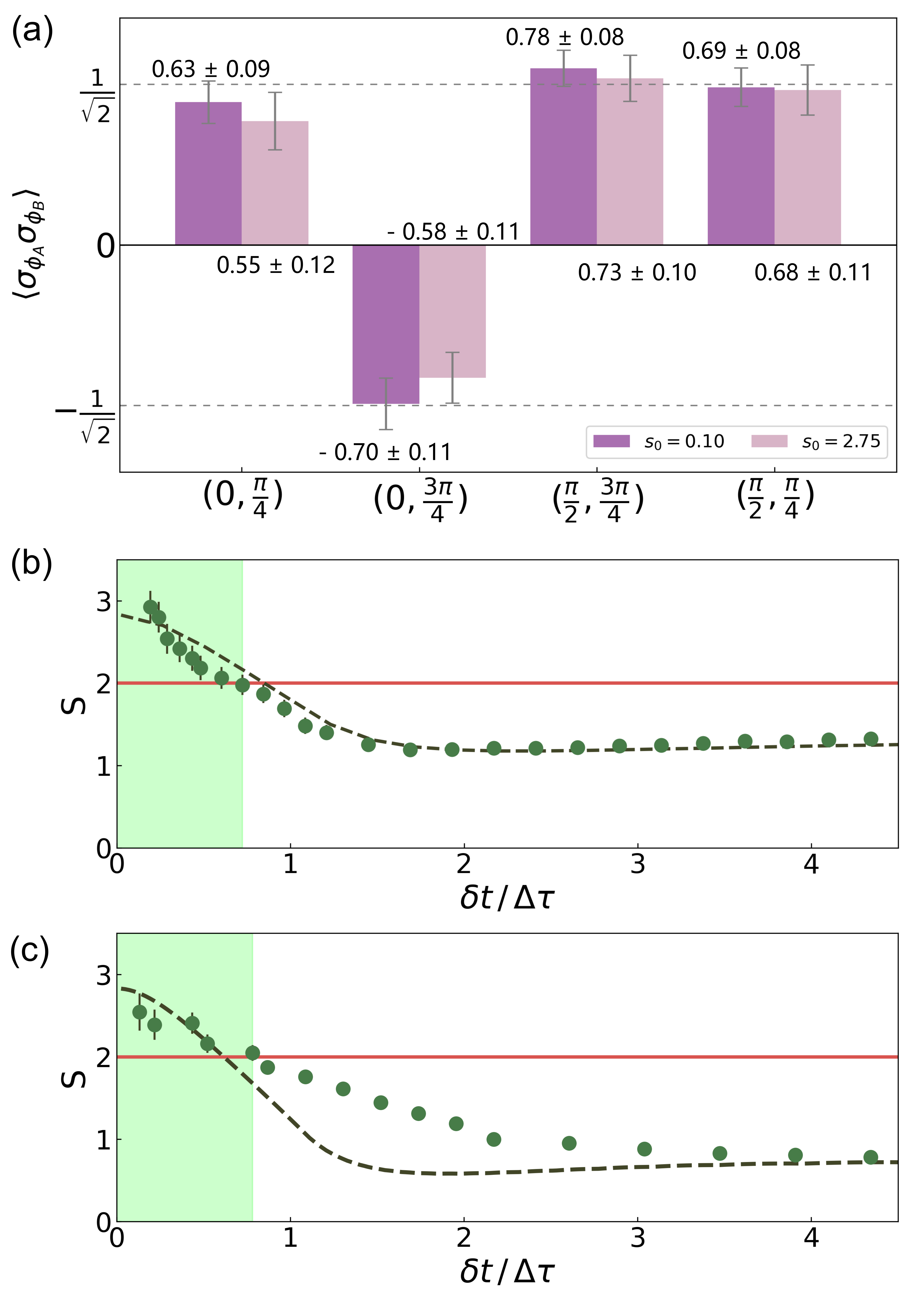}
\caption{\label{fig3} Violating the CHSH inequality with resonance fluorescence. (a) Measured two-photon correlations of the outputs of \emph{Alice's} and \emph{Bob's} interferometer for different settings of the phases $\phi_A$ and $\phi_B$. From these correlations, we deduce $S=2.80\pm 0.19$ for low saturation and $S=2.55\pm 0.22$ for high saturation. (b) and (c) Measured $S$-parameter (green dots) as a function of the coincidence time window $\delta t$ in units of the antibunching timescale $\Delta\tau$, see main text. The dashed line shows the theoretical prediction and the green-shaded area indicates where the measured $S$-parameters violate the CHSH inequality. For larger values of $\delta t/\Delta\tau$, $S$ approaches the characteristic level of uncorrelated photons.}
\end{figure}
\par In order to quantify the amount of entanglement in the photon pairs, we study the violation of a CHSH-type Bell inequality \cite{CHSH69} by measuring the $S$-parameter 
\begin{equation}\label{Sparameter}
S=\left|\langle\sigma_{\phi_A}\sigma_{\phi_B}\rangle-
\langle\sigma_{\phi_A}\sigma_{\phi_{B'}}\rangle+
\langle\sigma_{\phi_{A'}}\sigma_{\phi_B}\rangle+
\langle\sigma_{\phi_{A'}}\sigma_{\phi_{B'}}\rangle\right|.
\end{equation}
Using optimal phase settings $\phi_A=0$, $\phi_{A'}=\frac{\pi}{2}$, $\phi_B=\frac{\pi}{4}$ and $\phi_{B'}=\frac{3\pi}{4}$, the sum of the joint expectation values should reach $S=2\sqrt{2}\approx2.828$ for a maximally entangled state. Figure~\ref{fig3}(a) shows the measured two-photon expectation values for our phase settings under both low- and high-saturation regimes, using a coincidence window of $\delta t=\pm 10$~\unit{\ns} (for $s_0=0.10$) and $\delta t=\pm 3$~\unit{\ns} (for $s_0=2.75$) as depicted in Fig.~\ref{fig1}(b) \footnote{This time window was chosen by the condition that the theoretically expected second order correlation $g^{(2)}(\delta t)\approx 0.15$ and thus should yield the same quality of entanglement for both driving strengths}. We measure $S = 2.80 \pm 0.19$ and $S = 2.55 \pm 0.22$ in the low and high-saturation regimes, respectively. This corresponds to a $4.2\sigma$ and $2.5\sigma$ violation of the local realistic boundary of $S\leq2$ and clearly shows the high degree of entanglement of the measured photon pairs.

The degree of Bell violation—and thus the amount of entanglement—depends on the ratio between the coincidence window width, $\delta t$, and the characteristic antibunching timescale, $\Delta\tau$, defined by $g^{(2)}(\pm \Delta\tau)=1/2$. As $\delta t$ increases, photon antibunching is progressively averaged out, reducing the observed correlations. To quantify this effect, we measure the dependence of $S$ on $\delta t/\Delta\tau$ (see Fig.~\ref{fig3}). We observe $S>2$ for both driving strengths when $\delta t/\Delta\tau \lesssim 0.8$. Notably, the Bell violation occurs at similar values of this normalized ratio, even though $\Delta\tau$ itself differs significantly between weak and strong driving ($\Delta\tau = 41.5$~ns and $\Delta\tau = 23.0$~ns, respectively). This highlights the central role of the antibunching timescale in determining the observability of entanglement.

The coincidence window also sets the effective spectral resolution of the detection, $\Delta\omega \approx 1/\delta t$. For our experimental parameters, this corresponds to $\Delta\omega = 2\pi \times 5.3$ MHz (weak excitation) and $\Delta\omega = 2\pi \times 9.4$ MHz (strong excitation). These values are consistent with the spectral width of the incoherent component of the fluorescence, which is determined by the excited-state decay rate, $2\gamma = 2\pi \times 6$ MHz, in the weak-driving regime, and approximately by the Rabi frequency, $\Omega \approx 2\pi \times 10$ MHz, under strong driving.

Entanglement is therefore observed when the coherently and incoherently scattered photons are spectrally indistinguishable within the detection bandwidth, allowing for their interference. This indistinguishability is also a prerequisite for high-contrast photon antibunching \cite{dalibard83,hanschke2020,Masters2023}, emphasizing that quantum interference between the coherently and incoherently scattered multiphoton components underlies both phenomena. For higher spectral resolution (i.e., larger $\delta t$), the two contributions become progressively distinguishable, their interference is suppressed, and $S$ approaches the value expected for uncorrelated photons (see Supplementary Materials). The essential role of this interference distinguishes our work from approaches that isolate only the incoherent scattering contribution, either through spectral filtering \cite{Liu2024,Wang2025} or via fully incoherent excitation \cite{Trebbia2010,Rezai2019}. 
Notably, as the destructive interference prevails for all driving strengths, one expects no change of entanglement with increasing saturation.

\begin{figure}[ht]
\centering
    \includegraphics[width=0.45\textwidth,keepaspectratio]{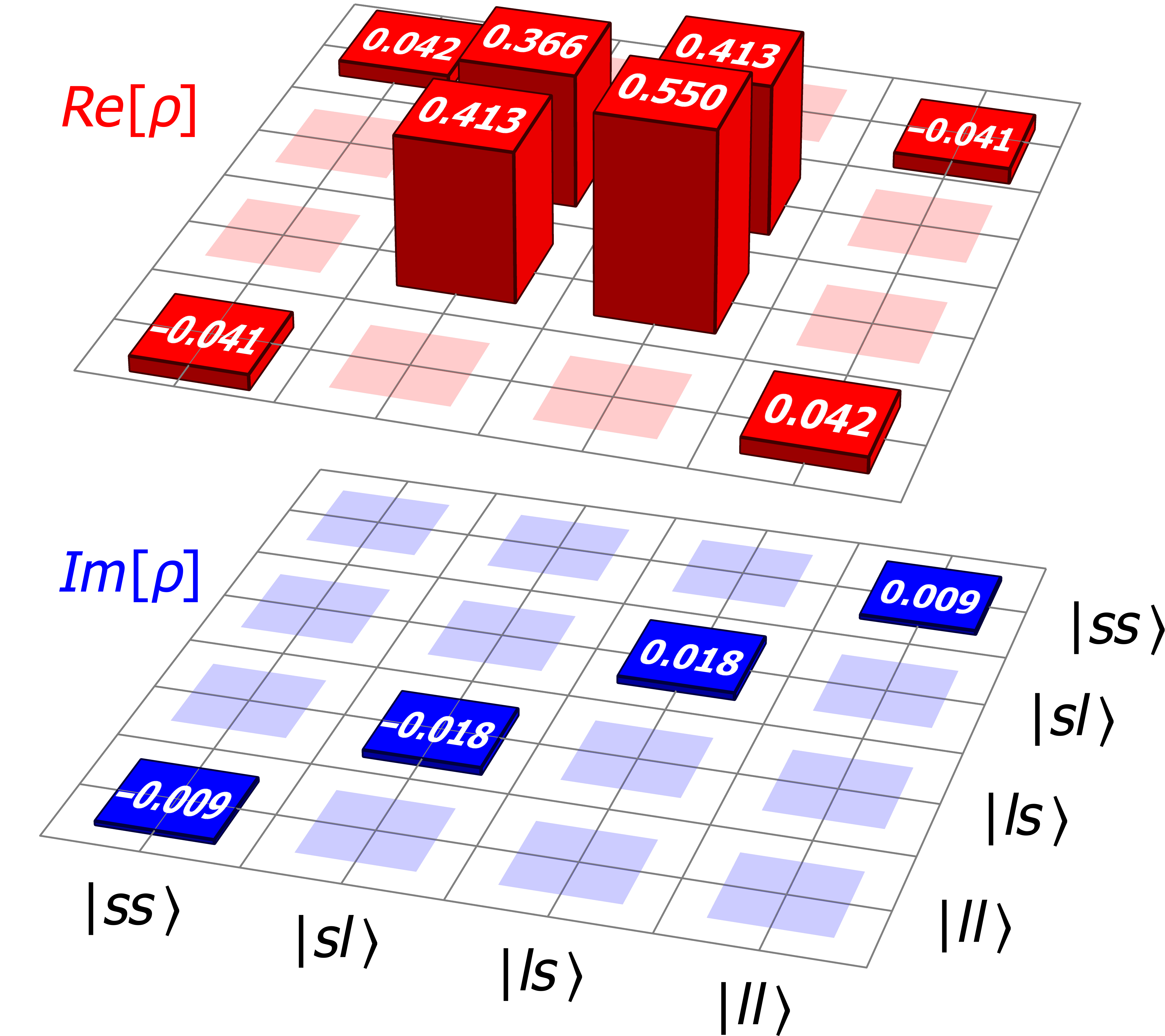}
\caption{\label{fig4}Tomographically reconstructed density matrix. The real (red) and imaginary (blue) parts of $\rho=\ket{\Psi_\text{Bell}}\!\bra{\Psi_\text{Bell}}$ measured in a $\delta t=\pm10$~\unit{\ns} coincidence window centred at zero time delay. Non-zero entries are labeled accordingly.}
\end{figure}
\par To further quantify the photon-photon entangled state, we perform a tomographic reconstruction of its density matrix $\rho$ for the weak excitation regime, see Supplementary Materials for details. Figure \ref{fig4} shows the reconstructed matrix for a $\delta t=\pm 10$~\unit{\ns} coincidence window. Calculating the overlap between the measured density matrix $\rho$ and the expected Bell state, we find a fidelity of $F=\langle \Psi_\text{Bell} | \rho | \Psi_\text{Bell} \rangle=0.87\pm0.02$, which again illustrates the high degree of entanglement of the two-photon state. Using the Peres–Horodecki criterion \cite{Horodecki1995} we can calculate the expected CHSH-inequality violation for the measured density matrix and obtain $S_F=2.57\pm0.05$, in agreement with the direct measurement of $S$ above.
The fact that we can extract such high-fidelity entangled states from resonance fluorescence by correlating the temporally filtered detection signals on Alice's and Bob's side shows that the emitted field is non-classical not only because of its antibunched photon statistics but also because of its underlying multimode entanglement.



Finally, it is worth noting that the resonance fluorescence can exhibit a high degree of first-order coherence, as evidenced by the high visibility of the interference fringes observed in the individual interferometers, see Fig.~\ref{fig2}. To understand how high-fidelity entangled photon pairs can emerge from such a coherent light field, it is necessary to examine the physical origins of both first-order coherence and entanglement in greater detail.
The high-visibility data in Fig.~\ref{fig2} arises from the coherently scattered component of resonance fluorescence, which dominates for weak saturation \cite{CohenTan2004}. In contrast, the coincidence detection rate---central to the entanglement measurement---is governed by the incoherent component, whose overall power contribution is negligible in this regime. As the saturation increases, the relative contribution of the coherent component decreases \cite{CohenTan2004}, leading to reduced visibility (Fig.~\ref{fig2}) but an enhanced rate of entangled photon pair generation. 
The maximum achievable pair rate is then determined by a trade-off between photon flux and the decreasing antibunching timescale $\Delta\tau$. Under optimal conditions, a maximum pair rate of approximately $7\,\%$ of the single-photon scattering rate $\gamma$ can be reached at $s_0 = 4$ (see Supplementary Materials).

\par In conclusion, we have experimentally demonstrated that entangled photon pairs can be extracted from the resonance fluorescence of a single quantum emitter. Our results highlight a fundamental connection between photon antibunching and photon–photon entanglement, both arising from quantum interference between the coherently and incoherently scattered components of the emitted light.
Notably, in our scheme, the photons are generated independently, and their entanglement originates from their indistinguishability combined with the intrinsic anticorrelations in the emission statistics of a two-level emitter. This distinguishes our approach from established entanglement-generation schemes---such as nonlinear parametric down-conversion \cite{kwiat1995,Valivarthi2016,Wang2016,Liscidini2023}, four-wave mixing \cite{Lin2007,Silverstone2014,Sharping2006,Harada2011}, or cascaded emission \cite{Dousse2010,Mueller2014,Trotta2016}---where photon pairs are produced simultaneously.
More generally, our findings indicate that any first-order coherent, antibunched light source \cite{prasad20,Cordier2023}  can be exploited to generate entangled photon pairs using this method. Finally, the presented scheme offers a pathway for extension to higher-dimensional entangled states, for instance through the analysis of three-photon coincidence events.

\begin{acknowledgments}
\par We acknowledge stimulating discussions with Anders S. S\o{}rensen and Martin Cordier and funding by the Alexander von Humboldt Foundation in the framework of the Alexander von Humboldt Professorship endowed by the Federal Ministry of Education and Research, as well as by the European Commission under the project DAALI (No.899275). X.-X. H. acknowledges a Humboldt Research Fellowship by the Alexander von Humboldt Foundation.

\par A.R. and J.V. conceived the experiment; X.-X.H., G.M. and L.M. designed the set-up; X.H., G.M. and L.M. performed the experimental work; X.-X.H., G.M. and J.V. developed the theoretical model; X.-X.H., G.M. and L.M. analysed the data; all authors contributed to the discussion of the results and the manuscript preparation.
\end{acknowledgments}

%
%
%
%
%
%
%

\clearpage
\onecolumngrid 
\begin{center}
    \textbf{\large Supplementary Materials for On the Relationship Between Antibunching and Entanglement in Resonance Fluorescence} \\
    \vspace{10pt}
    Xin-Xin Hu, Gabriele Maron, Luke Masters, Arno Rauschenbeutel$^*$, and J\"urgen Volz$^\dagger$ \\
    \textit{Department of Physics, Humboldt Universit\"at zu Berlin, 10099 Berlin, Germany}
    \small{Email: $^*$arno.rauschenbeutel@hu-berlin.de, $^\dagger$juergen.volz@hu-berlin.de}
\end{center}
\vspace{20pt}
\twocolumngrid 






\section{Theoretical model}\label{sect:entanglement}
In the following, we consider the case of continuous driving of the atom with a coherent light field. For a given time duration, we can expand the scattered light field in terms of its photon number components as
\begin{eqnarray}
\ket{\psi}&=&\ket{0}+\alpha\ket{1}+\beta\ket{2}+\ldots  \label{eqn:inputstate}
\end{eqnarray}
where $\ket{1}$ and $\ket{2}$ denote the number of photons in the scattered field and are given by 
\begin{eqnarray}
\ket{1}&=&\int dt\,a^\dagger(t)\ket{0}\\
	\ket{2}&=&\iint dt\, d\tau\,\psi(\tau)\,a^\dagger(t)\,a^\dagger(t+\tau)\,\ket{0}.\label{initial_state}
\end{eqnarray}
Here, $a^\dagger(t)$ is the operator for the creation of a photon at time $t$. Note that $\ket{2}$ is not a Fock state but contains temporal fluctuations that are described by the two-photon wave function $\psi(\tau)$, which in the case of weak driving can be explicitly written as \cite{Masters2023}
\begin{equation}
	\psi(\tau)=1-e^{-(\gamma-i\Delta)|\tau|},
\end{equation}
related to ${g^{(2)}(\tau)=|\psi(\tau)|^2}$ in the chosen time window.
After their generation and collection, the photons pass a first beamsplitter that directs them to \emph{Alice} and \emph{Bob}, as indicated by the operators $a^\dagger(t)$ and $b^\dagger(t)$, respectively. The corresponding two-photon state after the beam splitter is given by
\begin{equation}
\frac{1}{2}\iint dt\,d\tau\,\psi(\tau)\,a^\dagger(t)\,b^\dagger(t+\tau), \label{initial_state2}
\end{equation}
where the factor $1/2$ comes from the fact that we limit the discussion to the case where the two photons separate. The state in Eq.~(1) can be derived from this expression by accordingly choosing the times $t_1$ and $t_2=t_1-\tau$. At \emph{Alice} and \emph{Bob}, a second beamsplitter sends each photon into the short or long arm, introducing a propagation delay $\Delta t_A$ at \emph{Alice} and $\Delta t_B$ at \emph{Bob}, respectively. Assuming that both interferometers exhibit the same path length difference $\Delta t=\Delta t_A=\Delta t_B$, the state in Eq.~(\ref{initial_state2}) is transformed such that the integrand reads 
\begin{eqnarray}
&&\frac{1}{2}\psi(\tau)\Big[a_s^\dagger(t)b_s^\dagger (t+\tau)+a_s^\dagger(t)b_l^\dagger(t+\tau+\Delta t)\nonumber\\
&&+a_l^\dagger(t+\Delta t)b_s^\dagger (t+\tau)+a_l^\dagger(t+\Delta t)b_l^\dagger(t+\tau+\Delta t)\Big].\nonumber\\
\label{four_integrals}
\end{eqnarray}
Here, the subscripts $s$ and $l$ denote the short and long interferometer arms, respectively. As we assume an experiment with a continuous and coherent driving of the atom, the scattered photons are indistinguishable such that in Eq.~(\ref{four_integrals}) only time differences between photons matter. Consequently, the terms in the integrand can also be written as
\begin{eqnarray}
&&\frac{1}{2}\Big[\psi(\tau)a_s^\dagger(t)b_s^\dagger(t+\tau)+\psi(\tau-\Delta t)a_s^\dagger(t)b_l^\dagger (t+\tau)\nonumber\\
&&+\psi(\tau+\Delta t)a_l^\dagger(t)b_s^\dagger (t+\tau)+\psi(\tau)a_l^\dagger(t)b_l^\dagger(t+\tau)\Big].\nonumber\\
\label{final_state}
\end{eqnarray}

\par In the experiment, we are interested in the cases where \emph{Alice} and \emph{Bob} detect a photon coincidence with a small time delay $\delta t\ll\Delta t$, for which $\psi(\delta t)\approx0$. For large delay times $\Delta t \geq (2\gamma)^ {-1}$, the wavefunction does not vary much, so we can approximate $\psi(\Delta t +\tau)\approx\psi(\Delta t)$ and due to the time symmetry $\psi(\tau)=\psi(-\tau)$, Eq. (\ref{final_state}) simplifies to 
\begin{eqnarray}
&&\frac{1}{2}\psi(\Delta t)\Big[ a_s^\dagger(t)b_l^\dagger (t)+a_l^\dagger(t)b_s^\dagger (t)\Big]\nonumber\\
&&+\frac{1}{2}\psi(\delta t)\Big[ a_s^\dagger(t)b_s^\dagger (t)+a_l^\dagger(t)b_l^\dagger (t)\Big].\label{final_state2}
\end{eqnarray}
In the case of perfect antibunching and $\delta t\approx0$, only the first part in the above expression remains. We thus obtain the maximally entangled Bell state given by Eq.~(2) in the manuscript.

\par To measure the entanglement, we include a phase shifter that adds a phase shift $\phi_A$ and $\phi_B$ in the long arm of Alice's and Bob's interferometer, respectively, before closing them using another beamsplitter. Assigning the values $\pm1$ to the photon detections in the interferometer outputs, we can in this way measure the expectation values of the Pauli matrices $\langle\sigma_{\phi_{A(B)}}\rangle$ for the individual interferometers as well as the for the joint expectation values  $\langle\sigma_{\phi_A}\sigma_{\phi_B}\rangle$ where $\sigma_\phi=\cos\phi\,\sigma_x+\sin\phi\,\sigma_y$, see Supplementary Materials~\ref{chap:pauli}. Using $g^{(2)}(\tau)=|\psi(\tau)|^2$, we get for the expectation value of the joint measurement $\langle\sigma_{\phi_A}\sigma_{\phi_B}\rangle$ for two photons with time delay $\delta t$
\begin{eqnarray}
\langle\sigma_{\phi_A}\sigma_{\phi_B}\rangle&=&\frac{g^{(2)}(\Delta t)}{g^{(2)}(\delta t)+g^{(2)}(\Delta t)}\cos(\phi_A-\phi_B)\nonumber\\
&&+\frac{g^{(2)}(\delta t)}{g^{(2)}(\delta t)+g^{(2)}(\Delta t)}\cos(\phi_A+\phi_B).\nonumber\\
\label{eq:expectationvalue}
\end{eqnarray} 
From this expression, we get the limiting cases 
\begin{align}
\langle\sigma_{\phi_A}\sigma_{\phi_B}\rangle&\propto\cos(\phi_A-\phi_B) &\;\; (\delta t \approx 0 )\nonumber\\
	&\propto\cos\phi_A\cos\phi_B. &\;\; (\delta t \gg \Delta t )
\end{align}
Here, the first expression gives the expectation value for a maximally entangled state, while the second describes that of a fully separable state that is reached when the photons are detected with a large time delay. Equation~(\ref{eq:expectationvalue}) also shows that the degree of entanglement of the final state  is directly related to the quality of antibunching, i.e., $g^{(2)}(0)$, that can be reached in the experiment.
\par To quantify the strength of the observable entanglement signature, we consider the S-parameter of the CHSH-Bell inequality as defined in Eq.~(3). For the phase settings in the experiment, it reaches the maximum possible value $S_\text{max}$. Using Eq.~(\ref{eq:expectationvalue}), this maximum can be expressed as
\begin{equation}
    S_{\text{max}} = 2 \sqrt{2}\frac{g^{(2)}(\Delta t)}{g^{(2)}(\delta t) + g^{(2)}(\Delta t)}.
\label{eq:Smax_g2}
\end{equation}
The violation of the Bell inequality requires $S_\text{max} \ge 2$. Assuming sufficiently large delay times $\Delta t$ such that $g^{(2)}(\Delta t) \approx 1$, this allows one to derive a condition on the quality of the antibunching that is required to observe a Bell-inequality violation at all
\begin{equation}
    g^{(2)}(0) \leq \sqrt{2} - 1 \approx 0.414.
\label{eq:g2boundary}
\end{equation}

\par We note that throughout these derivations we made a weak driving approximation which allowed us to truncate the photon Hilbert space at two photons and to give an explicit expression for the two-photon wavefunction $\psi(\tau)$. However, for the case of sufficiently small time windows $\delta t$, we can always perform  the state expansion to up to 2nd order in photon number such that Eq.~(\ref{eq:expectationvalue}) also applies in strong driving cases as long as $\delta t$ is sufficiently small. Consequently, for $\delta t\approx 0$, Eq. (\ref{eq:Smax_g2}) applies for arbitrarily strong driving, thereby illustrating that the maximum entanglement fidelity does not depend on the driving strength.

\par In order to obtain a theoretical prediction of the influence of the size of the time window $\delta t$ on the value of $S$, one has to average Eq.~(\ref{eq:Smax_g2}) over all values of $\delta t$ in the coincidence window. The average value of $S$ that is obtained in this way is plotted as the black curve in Fig.~4. We note that for arbitrary $\delta t$, Eq.~(\ref{eq:Smax_g2}) is strictly only valid in the low driving regime, as our theoretical model does not provide an explicit expression similar to Eq.~(\ref{eq:Smax_g2}) for higher driving. In order to approximate the expected behavior, we assume that the entanglement will show a similar decrease of $S$ with increasing $\delta t$ as for small driving strengths. However, we take into account the high saturation by scaling the decay such that for large $\delta t$ it approaches the steady state value calculated for independent photons for the given driving strengths, see Supplementary Materials~\ref{sect:rate}.

\section{Measurement of photon states}\label{chap:pauli}

\par To measure the quantum state of the photons at \emph{Alice} and \emph{Bob}, we include a phase shifter that adds the phase shifts $\phi_A$ and $\phi_B$ in the long arm of each interferometer. Together with the final beamsplitter of the interferometer, this performs the transformation
\begin{eqnarray}
	c_s^\dagger&\rightarrow&\frac{1}{\sqrt{2}}(c_1'^\dagger+c_2'^\dagger),\\
	c_l^\dagger&\rightarrow&\frac{e^{i\phi}}{\sqrt{2}}(c_1'^\dagger-c_2'^\dagger),\label{transformations}
\end{eqnarray}
where $c_{s(l)}$ stands for $a_{s(l)}$ and $b_{s(l)}$, and the operator $c'_{1(2)}$ defines \emph{Alice's} and \emph{Bob's} output modes $a'_{1(2)}$ and $b'_{1(2)}$, respectively. Consequently, detecting a photon in the output $c'_{1,2}$ corresponds to detection of the state $(c_s^\dagger\pm e^{-i\phi}c_l^\dagger)/\sqrt{2}$, where $\phi\in\{\phi_A,\phi_B\}$. Assigning the values $\pm1$ to the detection events $c'_{1,2}$, this corresponds to a measurement of the Pauli matrices in the basis, $\langle\sigma_\phi\rangle=\langle\cos\phi\sigma_x+\sin\phi\sigma_y\rangle$, where $\sigma_x$ and $\sigma_y$ are the Pauli matrices in $x$- and $y$-direction.

\par Experimentally, the expectation values can be calculated  via
\begin{equation}
    \langle \sigma_{\phi_{A(B)}} \rangle = \frac{n_{a_1(b_1)}-n_{a_1(b_2)}}{n_{a_1(b_1)}+n_{a_1(b_2)}},
\end{equation}
where $n_{a_1(b_1)}$ ($n_{a_2(b_2)}$) is the number of detected photons in each output port. For the coincidence measurement between \emph{Alice} and \emph{Bob}, the joint expectation values can be calculated according to
\begin{equation}
   \langle \sigma_{\phi_A} \sigma_{\phi_B} \rangle=\frac{n_{a_1,b_1}+n_{a_2,b_2}-n_{a_1,b_2}-n_{a_2,b_1}}{n_{a_1,b_1}+n_{a_2,b_2}+n_{a_1,b_2}+n_{a_2,b_1}},
\end{equation}
where $n_{a_{1(2)},b_{1(2)}}$ is the number of detected coincidences between the different detectors.

\section{Coherence properties and rate of single photons}\label{sect:rate}
\par In our experiment, we observe simultaneously single photon coherence of the fluorescence light as illustrated in the interference pattern shown in Fig.~3, as well as coherence of the two-photon state which manifests itself as correlations between \emph{Alice} and \emph{Bob}. To get a better insight into these coherence properties, we analyze their dependence on the atom’s driving strength. In resonance fluorescence, the light scattered by a quantum emitter consists of two types of photons belonging to either a coherently or an incoherently scattered component, which are emitted by the atom with the respective rates
\begin{eqnarray}
    n_\textrm{coh}&=&\gamma\frac{s}{(s+1)^2},\\
    n_\textrm{inc}&=&\gamma\frac{s^2}{(s+1)^2},
\end{eqnarray}
where 
\begin{equation}
s=\frac{\Omega^2}{2\gamma^2+2\Delta^2}=\frac{s_0}{1+2(\Delta/\gamma)^2}
\end{equation}
with the on-resonance saturation parameter $s_0$ \cite{Steck07} and the detuning $\Delta$ between the atomic transition and the driving laser frequency. The total photon scattering rate of the atom is 
\begin{equation}
    n=n_\textrm{coh}+n_\textrm{inc}=\gamma \frac{s}{s+1}.
\end{equation}
\par As coherently emitted photons possess a well-defined frequency and phase given by the laser frequency $\omega_0$, they exhibit a well-defined interference fringe that, in principle, has perfect visibility. In contrast, incoherent photons exhibit a broad frequency distribution $\delta\omega\geq 2\gamma$ which is given by the Mollow-triplet \cite{Steck07}. For very long delays $\Delta t\gg (2\gamma)^{-1}$ in the unbalanced interferometer, the incoherent photons will thus not acquire a well-defined phase shift and consequently leave the interferometer with equal probability at each port, independent of the phase setting \footnote{For our experiment, the condition $\Delta t\gg (2\gamma)^{-1}$ is not completely fulfilled giving rise to a residual single photon visibility also for incoherent photons.}. 

\par In order to obtain a prediction of the single-photon visibility we have to calculate the photon rates in the interferometer outputs which are given by
\begin{equation}
    n_i(t)=\langle a_i^\dagger(t) a_i(t)\rangle\,,
\end{equation}
where $i=1,2$ labels the two different outputs. These output fields are the sum of two fields from the two interferometers via $a_{1(2)}^\dagger(t) = (a_s(t)\pm e^{i\phi} a_l(t))/\sqrt{2}$. As these operators can be expressed in turns of the atomic raising and lowering operators $\sigma^+$ and $\sigma^-$, respectively, the photon output rate is given by
\begin{eqnarray}
n_{1(2)}&=&\left\langle
\left[\sigma^+(t+\Delta t) \pm e^{-i\phi}\sigma^+(t)\right]\right.\nonumber\\
&&\left.\left[\sigma^-(t+\Delta t) \pm e^{i\phi}\sigma^-(t)\right]
\right\rangle.
\label{eq:input}
\end{eqnarray}

\begin{figure}
    \centering
    \includegraphics[width=0.7\linewidth]{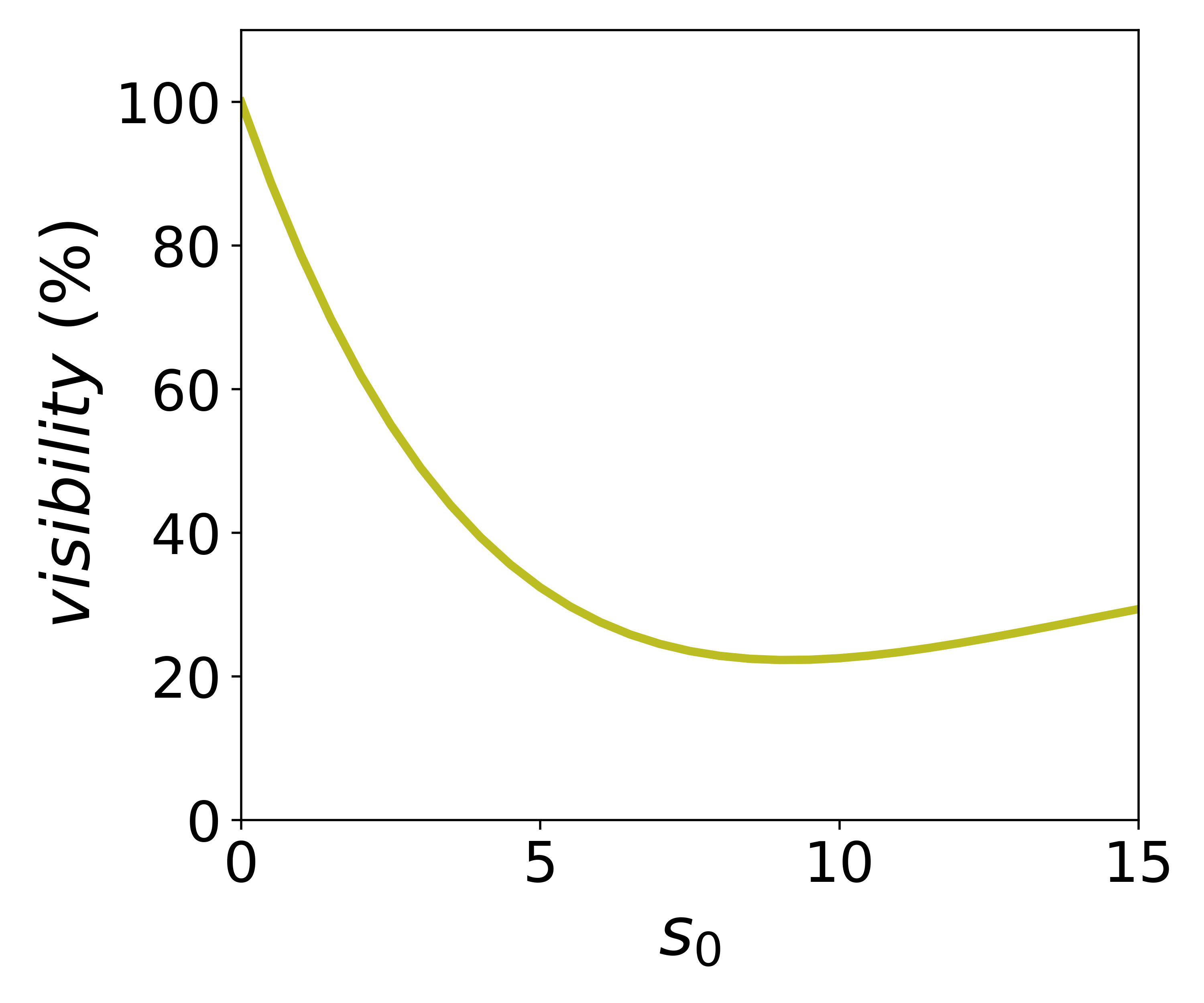}
    \caption{\label{figS1ext}\textbf{Theoretically calculated single-photon visibility versus saturation parameter.} Theoretical prediction of single-photon interference visibility $\text{V}$ as a function of the driving saturation parameter $s_0$, for our experimental parameters, see Supplementary Materials~\ref{sect:rate}.}
\end{figure}

Using the steady-state solutions from the optical Bloch equations together with the quantum regression theorem, the above expectation values can be calculated for our interferometer output modes. From these, we obtain for our experimental detuning of $\Delta = 2\pi\times 2.56$~\unit{MHz} a saturation-dependent visibility shown in Fig.~\ref{figS1ext}. In particular, for the driving strengths discussed in the manuscript, our theory predicts $\text{V}_\text{theory}=97.6\,\%$ for low saturation ($s_0=0.10$) and $\text{V}_\text{theory}=52.4\,\%$ for high saturation ($s_0=2.75$). We note that for high driving strengths we experimentally measure a higher visibility $\text{V}_\text{exp}=61.8\,\%$. This  discrepancy may originate from the different contributions of the individual experimental runs with fluctuating detuning to the averaged second-order correlation function (square of the scattered power) and the single photon interference fringes (linear in scattered power).

\par Using the above formalism also allows us to predict the two-photon expectation values for a large time delay $\delta t$. For $\delta t\to \infty$ the two-photon expectation simplifies to 
\begin{equation}
    \langle \sigma_{\phi_A}\sigma_{\phi_B} \rangle = \langle\sigma_{\phi_A}\rangle\langle\sigma_{\phi_B}\rangle.
\end{equation}

\section{Photon pair rate}
\par The rate of photon pairs that can be detected in a coincidence window $\pm\delta t$ is given by 
\begin{equation}
    n_p=2\left(\frac{n}{2}\right)^2\delta t=\frac{\gamma^2}{2}\frac{s^2}{(s+1)^2}\delta t.
\end{equation}
In the following we use the characteristic time scale of the antibunching for the coincidence window, i.e., $\delta t\approx=\gamma^{-1}$, with which the above expression simplifies to
\begin{equation}
   n_p=\frac{\gamma}{2}\frac{s^2}{(s+1)^2}=\frac{1}{2}n_\textrm{inc}. \label{eq:npair}
\end{equation}
This expression shows that the entangled photon pair events detected in our experiment are, up to a factor, equal to the rate of emission of the incoherently scattered photons, thus shedding light onto their physical origin. 

\par As the pair creation rate monotonously increases with the saturation of the atom, one good strategy to maximize the photon pair rate is to increase the atomic driving strength. However, stronger driving also narrows the antibunching interval in the second‐order correlation function requiring adjustment of the coincidence window to obtain a high fidelity entangled state. To quantify the effect of this trade-off, we analytically derive the dependence of the antibunching window on the driving power as follows.

\par We consider on-resonance driving for which the second-order correlation function is given by \cite{Steck07}
\begin{equation}
	g^{(2)}(\tau)=1-e^{-3\gamma\tau/2}\left(\cos\tilde\Omega\tau+\frac{3\gamma}{2\tilde\Omega}\sin\tilde\Omega\tau\right),
\end{equation}
with the Rabi frequency $\Omega$ and $\tilde\Omega^2=\Omega^2-\gamma^2$. For high driving strengths, this function will oscillate with the Rabi frequency, resulting in a shorter time window in which photon antibunching can be observed. To get an analytical expression for this time window, we approximate the second-order correlation function around $\tau=0$ by its Taylor expansion 
\begin{equation}
    g^{(2)}(\tau)\approx(2\gamma^2+\Omega^2)\tau^2+....
\end{equation}
from which we get the power-dependent width of the antibunching window of 
\begin{equation}\label{eq:timewindow}
\delta t=(\Omega^2/2+\gamma^2)^{-1/2}.
\end{equation}
Using this time window, together with Eq.~(\ref{eq:npair}), we obtain for the rate of entangled photon pairs 
\begin{equation}
n_p=\frac{\gamma^2\Omega^4}{4\sqrt{(\gamma^2+\Omega^2/2)^5}}.\label{eq:npair2}
\end{equation}
This expression has its maximum for $\Omega=2\sqrt{2}\gamma$ or $s_0=4$ and reaches a value of
\begin{equation}
   n_{p,max}= \frac{4}{25\sqrt{5}}\gamma\approx0.07\gamma,
\end{equation}
i.e., the maximum entangled photon pair rate reaches 7\,\% of the maximum possible single photon scattering rate $\gamma$.

\section{Trapping, detecting, and probing single atoms}\label{Expdetails}
We prepare a cloud of $^{85}$Rb atoms inside an ultra-high vacuum chamber using a magneto-optical trap (MOT), that is used as a reservoir of cold atoms for loading an optical dipole trap. The dipole trap is generated by focusing a laser beam (wavelength: $\lambda=784.65$~\unit{\nm}, waist radius: $w=1.334\pm 0.08$~\unit{\um}) into the MOT cloud using a high numerical aperture lens (AS-AHL12-10, Asphericon) (focal length: $f=10$~\unit{\mm}, working distance: $w_d=7.6$~\unit{\mm}) that is located inside the vacuum. Due to the microscopic trap volume, our trap operates in the collisional blockade regime \cite{Schlosser2001,Schlosser2002} such that, at most, a single atom is present inside the trapping volume at any time. For a laser power of $P=0.58$~\unit{\milli\W}, we obtain an optical trapping potential with a depth of $U/k_B=0.76$~\unit{\milli\K}, corresponding to trap frequencies of $\nu_r=66.8$~\unit{\kHz} and $\nu_z=9.1$~\unit{\kHz} in radial and axial directions, respectively. 

\par Resonance fluorescence photons originating from the trapping volume are collected with the same in-vacuum lens, separated from the trapping light using a dichroic mirror (LL01-785-25, Semrock), and coupled into a single-mode fiber that also acts as a spatial filter. Photons in the fiber are detected using superconducting nanowire single photon detectors, SNSPDs (Eos R12, Single Quantum), with each arrival time recorded by an FPGA-based timetagging unit. The presence of an atom inside the dipole trap is registered by an increase in the detected photon rate from the background level of $500$~\unit{\s^{-1}} to $3000$~\unit{\s^{-1}}. 

\par Following the detection of an atom in the dipole trap, an interleaved driving and cooling sequence is applied to the atom for a total duration of $200$~\unit{\ms}. In the driving interval, we send a driving laser beam onto the trap region, which is resonant to the light-shifted transition of the atom and is applied perpendicular to the trap axis in order to minimize stray light. After this probing, we apply the cooling laser of the MOT to cool the atom back to its initial temperature. Each $500$~\unit{\us} cycle comprises a $60$~\unit{\us} driving interval followed by $440$~\unit{\us} of cooling at low saturation, or a $3$~\unit{\us} driving interval followed by $497$~\unit{\us} of cooling at high saturation. A repumping field remains constantly on during the sequence. The duration of the driving and cooling times was optimized by maximizing the total rate of fluorescence photons detected during the driving process. For a low-saturation driving, we detect a photon rate of $2.51$~\unit{\kHz}, which agrees with the expected scattering rate under our low-excitation regime ($s_0=0.10$), when considering the limited collection efficiency of the lens and fiber ($\eta_0\approx1.2\;\%$), propagation losses through the Franson interferometer ($\eta_\textrm{prop.}\approx20\;\%$), as well as the average SNSPD detector efficiency ($\eta_\textrm{det.}\approx 86\;\%$). Proportionally, for a high-saturation driving, we detect a photon rate of $26.44$~\unit{\kHz}, showing a high-saturation driving strength with $s_0=2.75$.

\section{Fiber-based Franson interferometer}
Our Franson interferometer consists of two unbalanced Mach-Zehnder interferometers constructed using optical fibers spliced to commercially available 50:50 fiber beamsplitters (TN785R5A2, Thorlabs). 
The long arm of each interferometer includes a home-made piezo-based fiber-stretcher to control the interferometer phases $\phi_A$ and $\phi_B$. The whole setup is placed inside a thermally-insulated box with typical temperature stability of better than $0.1$~\unit{\degreeCelsius} on a daily time scale.
The optical path length difference between the long and short arms in each interferometer, $\Delta L_{A(B)}$, is measured by injecting a $\sim1$~\unit{\ns} duration pulse of light (wavelength: $\lambda\approx780$~\unit{\nm}) into the Franson interferometer and monitoring its arrival time on each output using the four SNSPDs.
From the observed delay time $\Delta t_A=46.1\pm0.2$~\unit{\ns} and $\Delta t_B=46.7\pm0.2$~\unit{\ns}, we calculate respective path length differences of $\Delta L_A=9.50\pm0.004$~\unit{\m} and $\Delta L_B=9.63\pm0.004$~\unit{\m}.

\par To ensure a polarization-independent operation of the interferometers, fiber birefringence is compensated by using in-line polarization controllers (CPC900, Thorlabs) in each interferometer to maximize the fringe visibility for orthogonal input polarizations. Following this procedure, we measure a visibility of $99\pm1\;\%$ in each interferometer averaged over four different polarizations of the input light (linear vertical and horizontal, left- and right-circular). The uncertainty in this estimation mainly comes from the background noise of the photodetectors.

\par During the experiment, each interferometer is set to impart a desired phase shift on the transmitted light. These phases are set by fixing the length of the long arm in each interferometer following a sample-and-hold locking procedure, which takes place every $30$~\unit{\s}. During a locking cycle, the driving light is injected into the Franson interferometer. An error signal is obtained by monitoring the difference in the count rate at the two outputs of each interferometer. To lock to the desired phase shift, a frequency shift is applied to the lock laser to move the zero-crossing of the error signal to the desired path length difference. Between locking cycles, the interferometer was free running with a maximal drift rate of the interferometer phase (likely due to slow thermal fluctuations) of about $2\pi\times0.022$~\unit{\radian\min^{-1}}. 

\section{Analysis of measured correlation data}
\par Anticorrelations in the light scattered by single atoms are measured using an HBT set-up, in which the output of the fluorescence collection fiber is connected to a 50:50 fiber beamsplitter with an SNSPD at each output (Fig.~2(b) in the main manuscript). Coincidence events were recorded in cycles following the same sequence as the CHSH measurement.
In addition, the data exhibits a slow bunching envelope on the microsecond timescale, arising from atomic heating during driving. 

\par To account for this, we fit the function $1+A\textrm{e}^{-|\tau|/t_b}$ to the coincidence data for large time delays. 
The value $1+A$ then serves as the baseline. After normalisation we observe $g^{(2)}(0)=0.05\pm0.036$ for $s_0=0.10$ and $g^{(2)}(0)=0.069\pm0.048$ for $s_0=2.75$.
The fits shown in Fig.~2(b) are based on the second-order correlation function of a two-level atom, taking into account a distribution of atom-light detunings to account for the different AC Stark shifts experienced by the atom due to its finite  temperature in the trap. The fit yields a mean residual detuning of the atomic resonance to the driving field of $\Delta=2\pi\times(2.56\pm0.16)$~\unit{\MHz}. 

\section{Maximum likelihood estimation}

\par The density matrix displayed in Fig.~5 in the main manuscript is reconstructed using a maximum likelihood estimation (MLE) method \cite{Banaszek1999, James2001}. For this, we define a physical density matrix $\rho$ as
\begin{equation}\label{eq:rho}
    \rho=\frac{T^{\dagger}T}{\mathrm{Tr}(T^{\dagger}T)}
\end{equation}
where $T$ is a $4 \times 4$ lower triangular complex matrix with free parameters that guarantee $\rho$ is positive, semidefinite and normalized. 
\par Each measurement setting, labeled by the indices $ij$, is described by a two-outcome positive operator-valued measure (POVM) with the projection operator
\begin{equation}
    P_{\pm}^{ij}=\frac{I\pm \langle \sigma_i \sigma_j \rangle}{2}.
\end{equation}
For a set of measurements $\{P^{ij}_+,P^{ij}_-\}$, we can define the likelihood function by
\begin{equation}\label{eq:likelihood}
    L(\rho)=\prod_{ij}\left[ \mathrm{Tr}(\rho P^{ij}_+)\right] ^{n^{ij}_+}\left[\mathrm{Tr}(\rho P^{ij}_-)\right]^{n^{ij}_-},
\end{equation}
where $n^{ij}_+=n_{a_1,b_1}+n_{a_2,b_2}$ and $n^{ij}_-=n_{a_1,b_2}+n_{a_2,b_1}$ are the number of detection events that yielded the outcomes $+1$ and $-1$ for the measurement $\langle \sigma_i \sigma_j \rangle$, respectively. 
Taking the natural logarithm of $L(\rho)$, we obtain
\begin{equation}
    \mathcal{L}(\rho)=\sum_{ij}\{n^{ij}_+ \ln{\left[\mathrm{Tr}(\rho P^{ij}_+)\right]}+n^{ij}_- \ln{\left[\mathrm{Tr}(\rho P^{ij}_-)\right]}\}.
\end{equation}

Introducing the total number of detected coincidences $N_{ij}=n^{ij}_++n^{ij}_i$ we rewrite the above equation and obtain the likelihood function as
\begin{eqnarray}
\mathcal{L}(\rho)&=&\sum_{ij}\frac{N_{ij}}{2}\{(1+\langle \sigma_i \sigma_j \rangle) \ln{\left[\mathrm{Tr}(\rho P^{ij}_+)\right]}\nonumber\\
&&+(1-\langle \sigma_i \sigma_j \rangle) \ln{\left[\mathrm{Tr}(\rho P^{ij}_-)\right]}\}.
\label{eq:mle}
\end{eqnarray}

\section{Reconstruction of the density matrix}

\par To reconstruct the density matrix,  we perform measurements with the phase settings $(\phi_A,\phi_B)=(0,0), (0,\frac{\pi}{2}), (\frac{\pi}{2},0)$ and $(\frac{\pi}{2},\frac{\pi}{2})$. The measured expectations and coincidences are summarized  in Table~\ref{tab1}. 

\begin{table}[ht]
\caption{\label{tab1}Measurement for density matrix reconstruction}
\begin{ruledtabular}
\begin{tabular}{ccc}
Projection& Expectations $\langle \sigma_i \sigma_j \rangle$ & Total coincidences $N_{ij}$\\\hline
${ \sigma_x \otimes \sigma_x }$ & 0.679 & 112 \\
${ \sigma_x \otimes \sigma_y }$ & 0.018 & 110 \\
${ \sigma_y \otimes \sigma_x }$ & 0.083 & 133 \\
${ \sigma_y \otimes \sigma_y }$ & 0.928 & 138\\
\end{tabular}
\end{ruledtabular}
\end{table}
\par To realize a full quantum tomography, measurements in all bases $\{ \sigma_i \otimes \sigma_j \}$ ($i,j = x, y, z$) should be performed. In our setup, direct measurements of expectation values involving the operator $\sigma_z$ are infeasible, since it would require disassembling the interferometer. Instead, we exploit the fact that the $z$-basis is the natural basis of the system. The expectation value $\langle \sigma_z \sigma_z \rangle$ can thus be determined from the probabilities of finding the photons in the four path configurations $|s,s\rangle$, $|s,l\rangle$, $|l,s\rangle$ and $|l,l\rangle$, where $|s\rangle$ and $|l\rangle$ represent photons in the short or long interferometer arms, respectively. As these probabilities are defined by the quality of the measured photon antibunching, we can extract the expectation value from evaluating the measured second-order correlation functions at time $\tau \in [-\delta t, \delta t]$ and at a time $\Delta t+\tau$, yielding

\begin{equation}
    \langle \sigma_z \sigma_z \rangle=-\frac{\int_{-\delta t}^{\delta t}d\tau\ g^{(2)}(\tau+\Delta t)}{ \int_{-\delta t}^{\delta t}d\tau\left( g^{(2)}(\tau+\Delta t)+g^{(2)}(\tau)\right)},
\end{equation}

where $\Delta t$ is the interferometer’s path delay. The correlations $\langle \sigma_z \sigma_x \rangle$ and $\langle \sigma_z \sigma_y \rangle$ cannot be obtained  in this way and are set to zero in our reconstruction, as they are absent in the expected state and their value has no impact on the fidelity of the reconstructed density matrix.

\par The density matrix $\rho$ in Fig.~5 is reconstructed by minimizing the negative log-likelihood $-\mathcal{L}(\rho)$ in Eq.~(\ref{eq:mle}), with $\rho$ parameterized as Eq.~(\ref{eq:rho}) to ensure physicality. Numerical optimization is performed via Mathematica's \texttt{FindMinimum} with the \texttt{QuasiNewton} method. For the fidelity of the generated Bell state we obtain $F = \langle \Psi_\text{Bell} | \rho | \Psi_\text{Bell} \rangle = 0.87\pm0.02$. Here, the statistical error is determined using a Bootstrap method, where we add Poissonian noise to the measured coincidences and followed by the density matrix reconstruction. For this, we generate a set of $100$ random density matrices and use the resulting standard deviations as error estimation.

\bibliography{apssamp}

\end{document}